\documentclass[preprint,12pt]{elsarticle}

\usepackage{graphics}
\usepackage{amssymb}

\journal{Acta Astronautica}

\begin{document}

\begin{frontmatter}

\title{A novel SETI strategy targeting the solar focal regions of the most nearby stars}

\author{Micha\"el Gillon}
\ead{michael.gillon@ulg.ac.be}
\address{Institut d'Astrophysique et de G\'eophysique,  Universit\'e de Li\`ege,  All\'ee du 6 Ao\^ut 17,  B\^at.  B5C, 4000 Li\`ege, Belgium}

\begin{abstract}
Many hypotheses have been raised to explain the famous Fermi paradox.  One of them is
that self-replicating probes could have explored the whole Galaxy, including our Solar System,
and  that they are still to be detected. In this scenario, it is proposed here that probes from 
neighboring stellar systems could use the stars they orbit as gravitational lenses 
to communicate efficiently with each other. Under this hypothesis, a novel SETI approach 
would be to monitor the solar focal regions of the most nearby stars to search for communication 
devices. The envisioned devices are probably not detectable by  imagery or stellar 
occultation, but  an intensive  multi-spectral monitoring campaign could possibly detect 
some communication leakages. Another and more direct option would be to message the 
focal regions of nearby stars in an attempt to initiate a reaction.
\end{abstract}

\end{frontmatter}

\section{Introduction}
\label{}
In 1959, two scientists proposed to search for interstellar communications from extrasolar civilizations
\cite{1}. The first application of this idea was performed a few months later \cite{2}, effectively inaugurating 
the modern era of SETI,  the Search for Extra-Terrestrial Intelligence. Since then, a large number of 
SETI projects have been conducted, most of them focusing on the search for radio emissions from 
advanced civilizations trying to communicate with us \cite{3}. 

Despite the fact that no confirmed extraterrestrial signal has been detected so far, many discoveries 
obtained within the last  fifty years have strongly reinforced the scientific rationale of the SETI enterprise. 
First, we now know that planets  are the normal by-products of stellar formation, and  that most stars of our 
Galaxy harbor planets, including terrestrial planets similar to Earth and orbiting in the so-called 
circumstellar habitable zone \cite{4, 5, 6}. Second, there is now strong evidence that life appeared on 
Earth a few hundred millions years at most after our planet  had cooled enough to support liquid water 
on its surface \cite{7}. This early emergence  is consistent with a  high probability of abiogenesis 
providing a few conditions are met, mostly liquid water on the surface of an organics-rich terrestrial planet 
(or satellite) with a stable source of energy \cite{8}. Third, biologists have discovered life forms surviving 
in some of the harshest environments of Earth. The majority of  these  so-called extremophiles 
are microbes (Eubacteria and Archaea), but some multicellular animal examples are also known 
(e.g.  Tardigrades, Antarctic krills). Extremophiles have revolutionized our vision of the adaptability 
 of life, significantly expanding the plausible limits for life on other planets \cite{9}. 

Taken as a whole, these results are consistent with the ubiquitous presence of life in the Galaxy.
Furthermore, many examples of evolutionary convergence known on Earth indicate 
that our high level of intelligence and its resulting technology could be not an extraordinary
stroke of luck, but one among the finite number of outcomes to the natural selection 
mechanisms that should be in work in any biosphere \cite{10, 11, 12}. All these considerations make 
the emergence of other civilizations in the Galaxy since its birth more than 10 Gy ago a plausible 
hypothesis that deserves to be comprehensively  addressed. 

Our own technological civilization is less than two hundred years old, and  we have already sent 
robotic probes to a large number of bodies of our Solar System. Our technology is certainly not 
yet mature enough to build a probe able  to reach one of the nearest stars in a decent time (i.e. within a few 
decades), but nothing in our physical theories precludes such a project. On the contrary, the constant 
progress in the fields of space  exploration, nanotechnology, robotics and electronics, combined with 
the development of new possible energy sources like fusion reactors or solar sails,  indicate that 
interstellar exploration could become a technological possibility in the future, provided that our 
civilization persists long enough. So, if extrasolar technological civilizations have existed in the Milky Way, 
we cannot discard the intriguing possibility that at least one of them has visited our Solar System. 
The plausibility of this hypothesis is strongly increased when considering the concept of self-replicating 
interstellar probes \cite{13, 14}, i.e. probes that would be sent by a civilization to a few neighboring
planetary systems where they would mine raw materials to create replicas of themselves that would 
head towards other nearby systems, replicate there, and so on. With such a strategy very similar to the 
growth of a bacterial colony, it has been shown that a single self-replicating probe based on conventional physics 
and existing technological concepts could spread throughout the Galaxy within a few hundreds of 
millions years at most \cite{14, 15}.

The absence of such `Von Neumann probes' in our Solar System has been presented as a strong argument
against the existence of extraterrestrial civilizations \cite{14}. This new formulation of the 
Fermi paradox  \cite{16} assumes that an extraterrestrial probe monitoring the Solar System would 
be easy to detect. This is far from the truth. From the Sun to the external limit of the Oort cloud, the Solar System
 is a $5\times10^{14}$ AU$^3$ volume that remains mostly unexplored  \cite{17, 18}. Several unsuccessful 
 attempts have been made to detect extraterrestrial artifacts in the Solar System (e.g. \cite{19, 20, 21}), 
 but they have probed only a tiny  fraction of the Solar System, and concluding  at this  point that our Solar 
 System does not contain any extraterrestrial probe is definitely premature. Here we present a new strategy 
 for testing the existence of  such probes, based on a hypothesis presented in the next section.

\section{Proposed hypothesis}

The concept of Von Neumann probes exploring the Galaxy requires some level of communication 
and coordination between the probes. One could hypothesize that the probes would directly 
communicate with their original system, but this hypothesis is {\it a priori} very unlikely.
The vastness and structure of the Milky Way makes impossible (at least for our  technology) 
a direct communication between a large fraction of stars, for instance between the Sun 
and a star located at the opposite side of the  galaxy and hidden by the galactic center. 
Furthermore, the coordination of probes to explore neighboring systems would be very inefficient, 
considering the limited speed of light and the large distances in play.
A communication strategy based on direct connexions between neighboring systems would
be a much better solution, with the extra-benefit that the information gathered by probes would 
be spread among their whole network, without any loss even in case of collapse or migration of
 the original civilization. The first part of our hypothesis is thus that the envisioned
 probes would use this direct communication strategy.
 
The second part of our hypothesis concerns the technique used by the  probes for their 
interstellar communication. In this respect, every star represents a huge advantage, thanks to 
its large mass that, according to Einstein's theory of general relativity \cite{22}, deflects the light 
rays passing close to it. This effect, observed for the first time for the Sun in 1919 \cite{23}, makes 
any star a potential gravitational lens. As first outlined by Von Eshleman in 1979 \cite{24}, using the 
Sun as a gravitational lens would provide an unprecedented potential for astronomical observations 
and interstellar communication. This potential was further studied by Maccone \cite{25,26,27,28}
who confirmed the tremendous advantage of the effect for efficient interstellar communication.  For 
instance, at 32 GHz, the combined transmission  gain brought by the Sun and Alpha Cen A is 
$\sim$10$^{16}$, making possible an efficient communication between the two stars with very
 moderate transmission powers. In fact, Maccone's most striking conclusion is that it is {\it only} 
by using the gravitational focusing potential of the Sun that we would be able to communicate with a 
probe sent to any of the nearest stars \cite{29}.  Basing on this conclusion, it is assumed here that  
Von Neumann probes would use the same technique. This is the second part of our hypothesis. 

\section{Testing the hypothesis}

Assuming the colonization of the whole Galaxy by self-replicating probes, our working hypothesis presented 
above predicts that we should find an interstellar communication devices (ICD) at the solar focus of at least 
one nearby star.  This `focus' is in fact not localized, but any point above a specific distance on a line connecting
 the star to the Sun, in the opposite direction to the star. Using the Schwarzschild metric to represent the 
 gravitational field around the Sun, one can compute a value of 550 AU for this  minimum distance
  (e.g. \cite{28}). In practice, a larger distance is required, especially at lower frequencies, because of the 
  scattering properties of the  solar corona. A typical distance of 1000 AU is probably a better 
  estimation \cite{28}. 

For illustrative purpose, we consider from now on Alpha Centauri as the target system. From Earth, the position of the 
putative ICD relative to the background stars would not be fixed but would show a large amplitude 
(a few arcmin) 1yr-period elliptical movement (in first approximation) due to the annual parallax (Fig.~1). 
This corresponds to a mean offset of 5'' per day. In addition, the ICD would show a  motion in opposite direction to
the proper motion of Alpha Cen (3.6''/yr in RA and 0.7''/yr in DEC). 

To assess the potential for detecting the envisioned ICS, some working assumptions are now made about it, 
based on our current or foreseen technology. For the antenna itself, a diameter of one or two dozens of meters
at most can be assumed, based on the design of FOCAL, a proposed space mission that would use
the Sun as a gravitational lens \cite{25,26,27,28,29,30}. That is the whole point of using a star as lens for interstellar
 communication: relatively small antennae and emission powers are required. For the mass of the payload, 
 it is assumed to be not much larger than 1 ton. As a comparison, the mass of the Voyager 1 spacecraft is 
 722 kg.  By keeping the same azimuthal position in a barycentric inertial system, the ICD would drift radially 
 towards the Sun. To compensate for the gravitational pull of the Sun, a solar sail solution seems promising as 
 it requires no intrinsic energy source, so it is adopted here for our ICD toy model. A circular sail facing the Sun 
 is assumed, adopting as effective radiation pressure at 1AU $P_{eff}$ = 9.1$\mu$N$/$m$^2$, corresponding 
 to an ideal sail (perfectly flat, 100\% specular reflection) \cite{31}. For the sail surface density $\rho_s$, an upper
  limit is imposed from the following equation that equals the gravity and radiation forces:
\begin{equation}
\frac{G M_\ast (M_p+S_s \rho_s)}{r^2} = P_{eff} S_s
\end{equation} 
where $G$ is the gravity constant, $M_\ast$ and $M_p$ are respectively, the mass of the Sun and of the payload, 
$S_s$ is the surface of the sail, and $r$ = 1 AU. Rearranging
to solve for the surface of the sail leads to
\begin{equation}
S_s = \frac{G M_\ast M_p}{r^2 P_{eff} - G M_\ast \rho_s}
\end{equation} With $P_{eff}$ = 9.1$\mu$N$/$m$^2$, this equation is physically valid only for $\rho_s < 1.4$ g/m$^2$, making this 
value an absolute upper limit for the sail surface density. Solar sail material of such low density has not yet been manufactured on an industrial 
scale, but still prototypes composed of nanotubes and with densities down to 0.1g/m$^2$ do exist \cite{32}. For the purpose of this work, a 
 value of 0.5 g$/$m$^2$ is adopted. This results into a circular sail of $\sim$550 meters radius and of $\sim$ 0.5 ton mass. 

 Figure 2 shows the visual magnitude of the sail at zero phase angle as a function of the
 payload mass, assuming a Bond albedo of 1 and neglecting light scattering. For a 1 ton payload, the
 visual magnitude would be $\sim$ 30.5. This is in fact an absolute lower limit, as for an actual sail, the 
 reflection would not  be purely specular. Furthermore, assuming a zero phase angle means that 
 the telescope would have to be on the line connecting the ICD to the Sun, i.e. it would have to transit 
 the Sun as seen from the ICD. For an Earth-bound telescope and  an ICD targeting Alpha Cen,
 this is not possible, as Alpha Cen is not in the ecliptic plane (ecliptic latitude = 43$^\circ$). 
 Even assuming one could observe the ICD sail at zero phase angle, the use of the exposure time calculator
 simulator for the future E-ELT 39\,m telescope \cite{33, 34} results in  a cumulative exposure time of several
  weeks to detect a 30.5-magnitude point-source with a decent signal-to-noise ratio of 10. This means that 
  detecting the envisioned ICD using direct imaging techniques is not feasible with existing or upcoming facilities.
 
The stellar occultation technique has been presented as a promising method for detecting small objects
of the outer Solar System \cite{35}. Using the formalism presented in \cite{36}, we estimated the parameters 
of the occultation pattern for a faint and distant background star occulted by the envisioned ICD, assuming 
observations at 600 nm. The angular size of the sail at 1000 AU would be $\sim$ 1.5 $\mu$arcsec, making 
the occultation pattern completely dominated by diffraction. The angular radius of the occulted star was assumed 
to be 20 $\mu$arcsec, corresponding to a $\sim$ 2 $R_\odot$ star at 500 pc. This assumption does not alter the 
conclusions. The result is an occultation of $\sim$ 7 \% depth at most (for a null impact parameter) and $\sim$ 1 s 
duration. Detecting such a short and shallow photometric event on a (most probably) very faint star would be a 
real challenge with existing or planned facilities. Furthermore, the expected number of occultations per year is 
only $3.5 \times 10^{-4}$, basing on the field density of stars brighter than magnitude 21 estimated from the GSC-II 
and USNO catalogues, and on the sky area covered by the ICD during its parallactic ellipse. Thanks to the proper 
motion of Alpha Cen, the sail  would probe a slightly different area of the sky each year, nevertheless the  expected time 
between two occultations is of the order of several thousand years, leading to the conclusion that  the 
occultation technique is not a suitable method of detection either.

The two remote observational techniques commonly used to detect faint and small objects in the Solar System 
should thus be unable to detect the envisioned ICD. Another option is of course {\it in situ} exploration.  In case of 
realization of a FOCAL-like space mission targeting Alpha Cen or another nearby star in the future, it would be 
interesting to equip the spacecraft with instruments able to explore efficiently the focal region in search of the 
envisioned ICD,  not only efficient imaging  devices but also powerful radar and/or lidar systems.

Attempts to detect the hypothesized ICD can still be performed now, basing on the very purpose of the device: not only to 
receive messages from Alpha Cen, but also to send messages to Alpha Cen and to one or several probes orbiting the Sun. 
An intense multi-spectral monitoring of the focal region of Alpha Cen with, e.g., the Allen Telescope Array \cite{37}, 
could in principle detect some leakages in these communications, depending on the used technology, communication frequency, 
and emission power. 

Last but not least, a more direct detection strategy exists. Under the presented hypothesis, the astronomical position of the 
ICD is very well known, so we could simply attempt to message it, with the hope  to cause its reaction (either by responding 
back, or via an increased chatter with other probes).  Such an `active SETI'  \cite{38}  strategy has been criticized by 
several scientists basing on the risk of revealing our location to an hostile expansionary civilization (e.g. \cite{39}, \cite{40}). 
In the present case, such criticisms would not apply, as it is difficult to imagine how extraterrestrial robotic probes having 
monitored our Solar System for probably a very long time could not already know that Earth does host life, including our nascent 
civilization. 

\section{Conclusion}
It was hypothesized here that our Galaxy has been colonized by self-reproducing probes and that probes in neighboring 
systems communicate regularly using the stars they orbit as gravitational lenses for their interstellar
 communication. Under this hypothesis, an ICD should be present in the Solar System, in the focal region of (at least) 
 one nearby star, at $\sim$ 1000 AU for the Sun. It was argued that the traditional techniques of optical imaging and stellar 
 occultation would probably not be able to detect the envisioned ICD, and that our best chances of detection in the near future 
 relies on the intense multi-spectral monitoring of these regions, with the hope to catch a communication leakage, 
 and on directly messaging the focal regions in attempts to initiate a reaction of the envisioned 
 probes. While negative results could be explained not only by the non-existence of the envisioned ICD but also by a 
 stealth and discretion policy of the hypothesized probes,  a positive result would definitely revolutionize our understanding 
 of our place in the Universe. 

\section*{Acknowledgments}

The author is a Research Associate of the Fonds National de la Recherche Scientifique-FNRS., and  thanks
 B.-O. Demory, L.~Delrez, G. Harp, E. Jehin, C. Maccone, P. Magain, J. C. Tarter, A. H. M. J. Triaud \& A. Zaitsev for useful 
 comments and suggestions.

\newpage

\begin{figure}[t]
\label{fig:1}
\centering                     
\includegraphics[width=14cm]{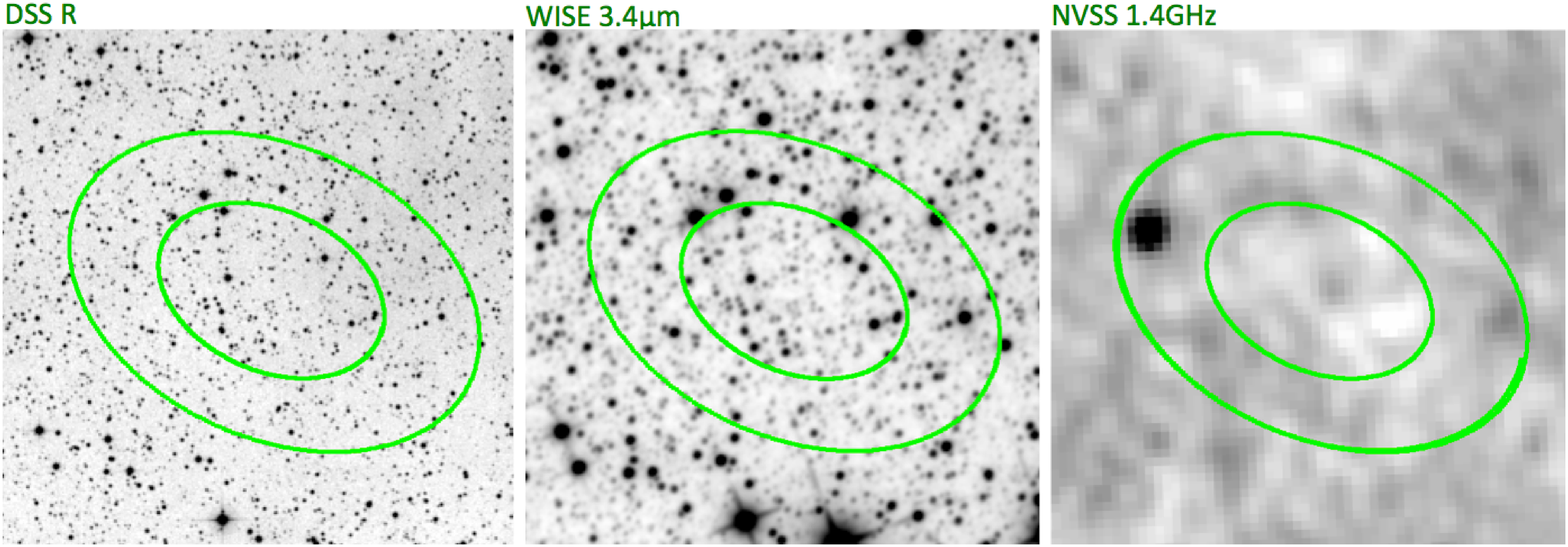}
\caption{DSS R-band (left), WISE-1 (middle) and NVSS (right) 15'$\times$15' image centered on the equatorial
coordinates opposite to the ones of Alpha Cen. For each image, the parallactic ellipse of the envisioned ICD is 
shown for distance to the Sun of 550 AU (outer) and 1000 AU (inner).}
\end{figure}

\begin{figure}[b]
\label{fig:2}
\centering                     
\includegraphics[width=12cm]{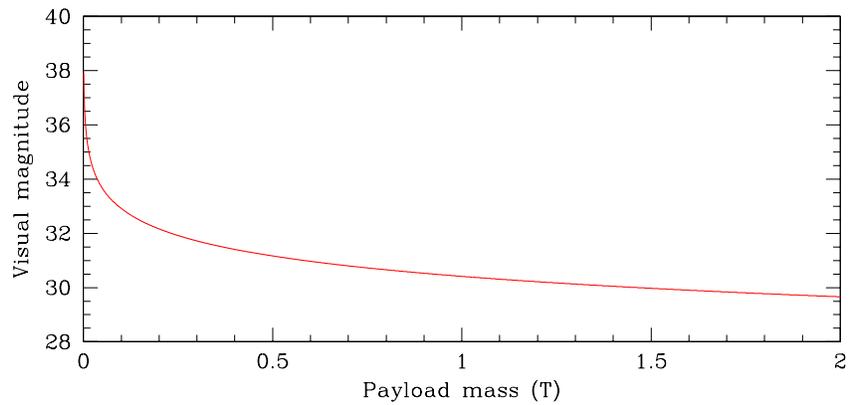}
\caption{Estimation of the optical magnitude of the ICD as a function of the payload mass. An ideal plane sail
observed at zero phase angle is assumed. }
\end{figure}

\end{document}